# Dirac mass generation from crystal symmetry breaking on the surfaces of topological crystalline insulators

Ilija Zeljkovic[*1], Yoshinori Okada[*1,2], Maksym Serbyn[3], R. Sankar[4], Daniel Walkup[1], Wenwen Zhou[1], Junwei Liu[3], Guoqing Chang[5], Yung Jui Wang[6], M. Zahid Hasan[7], Fangcheng Chou[4], Hsin Lin[5], Arun Bansil[6], Liang Fu[3] and Vidya Madhavan[1]

[1]Department of Physics, Boston College, Chestnut Hill, Massachusetts 02467, USA [2]WPI-AIMR, Tohoku University, Sendai, 980-8577, Japan [3]Massachusetts Institute of Technology, Cambridge, MA 02139, USA, [4]Center for Condensed Matter Sciences, National Taiwan University, Taipei 10617, Taiwan [5]Graphene Research Centre and Department of Physics, National University of Singapore, Singapore 117542 [6]Department of Physics, Northeastern University, Boston, Massachusetts 02115, USA [7]Joseph Henry Laboratory, Department of Physics, Princeton University, Princeton, New Jersey 08544, USA

*These authors contributed equally to this work

**The tunability of topological surface states (SS) and controllable opening of the Dirac gap are of fundamental and practical interest in the field of topological materials. In topological crystalline insulators (TCIs), a spontaneously generated Dirac gap was recently observed, which was ascribed to broken cubic crystal symmetry. However, this structural distortion has not been directly observed so far, and the microscopic mechanism of Dirac gap opening via crystal symmetry breaking remains elusive. In this work, we present scanning tunneling microscopy (STM) measurements of a TCI $Pb_{1-x}Sn_xSe$ for a wide range of alloy compositions spanning the topological and non-topological regimes. STM topographies directly reveal a symmetry-breaking distortion on the surface, which imparts mass to the otherwise massless Dirac electrons — a mechanism analogous to the long sought-after Higgs mechanism in particle physics. Remarkably, our measurements show that the Dirac gap scales with alloy composition, while the magnitude of the distortion remains nearly constant. Based on theoretical calculations, we find the Dirac mass is controlled by the composition-dependent SS penetration depth, which determines the weight of SS in the distorted region that is confined to the surface. Finally, we discover the existence of SS in the non-topological regime, which have the characteristics of gapped, double-branched Dirac fermions.**

TCIs belong to a distinct category of topological materials with mirror Chern number ±2 [1]. The topological SS in the (001) plane consist of two pairs of Dirac nodes, located in the vicinity of $\bar{X}$ and $\bar{Y}$ points in the two-dimensional Brillouin zone (BZ) [1]. In contrast to Dirac SS in three-dimensional topological insulators [2–6] which are protected by time-reversal symmetry [7–9], each pair of Dirac nodes in TCIs is protected by its respective mirror symmetry plane [10]. Based on earlier Landau level (LL) spectroscopy, it has been hypothesized [11] that in the topological state of $Pb_{1-x}Sn_xSe$, mirror symmetry is spontaneously broken in one of the two directions, thereby providing mass to one pair of Dirac nodes while leaving the other pair massless. However, the structural distortion postulated to do this has not yet been directly observed. In this work, we use scanning tunneling microscopy and spectroscopy to track the SS evolution in a TCI [1,10], $Pb_{1-x}Sn_xSe$ [12] that can be tuned continuously by changing the alloy composition ($x$) from the topological to the trivial phase [13]. By obtaining high-resolution topographies of the surface structure, we have succeeded in imaging the mirror symmetry breaking distortion. By simultaneously performing LL spectroscopy[14–17], we track the evolution of the mass as well as the distortion across the phase diagram. Surprisingly, we find that while the mass decreases monotonically as we approach the phase transition and vanishes at the critical point ($x_c$), the distortion magnitude remains relatively unchanged across the phase diagram. This is inconsistent with the simple picture that the changing mass results from a varying distortion. Instead, our observations can be explained by the increasing penetration of the SS into the bulk as we approach $x_c$. In this scenario, the distortion observed in STM is restricted to the surface. The magnitude of the mass is therefore controlled by the weight of the SS at the topmost layer, which decreases with the increasing bulk penetration of the SS. Our results present direct experimental evidence in support of the theoretical picture of the fate of the SS electrons close to the trivial composition where the trivial state is achieved via a complete penetration of the bulk of the sample by the Dirac electrons. Finally, on the trivial side, we discover novel SS, which surprisingly exhibit many of the characteristics of their topological counterparts.

$Pb_{1-x}Sn_xSe$ single crystals in the range of Sn concentrations studied in this work are expected to have a face-centered cubic structure[18] (Fig. 1). From the square atomic lattice seen in scanning tunneling microscopy (STM) topographs (Figs. 2(c,d)), the crystals cleave along the (001) direction. To explore the SS evolution on the (001) face, we perform LL spectroscopy in magnetic field on a series of samples spanning the quantum phase transition from the trivial to the topological phase: *x ~ 0.09 (trivial), 0.17 (critical), 0.29 (topological) and 0.34 (topological)* (Fig.3). Our first task is to ascertain the existence of Dirac SS in the topological samples. In order to accomplish this, we determine the existence of the Dirac nodes at different Sn concentrations across the transition point (Fig. 3). The Dirac node signifies the continuous nature of the topological SS bands, which necessarily span the gap between the conduction and valance bands. The presence of this topologically protected point is manifested as a non-dispersing (peak energy independent of magnetic field) $0^{th}$ LL peak in *dI/dV* spectra located near the minimum. We find that such a non-dispersing peak is clearly seen in all three non-trivial samples (Figs. 3(f-h)). In contrast, although LLs signifying the existence of SS are clearly observed even in the trivial *x*~0.09 sample (Fig. 3(e)), the $0^{th}$ LL is notably absent, which indicates the non-topological nature of the SS.

In the topological samples, in addition to the peak at the Dirac point, we find two additional non-dispersing peaks labeled $E^*_+$ and $E^*_-$ (Figs. 3(c,d,g,h)) on either side of the $0^{th}$ LL peak. From an earlier study [11], these peaks correspond to mass acquisition in two out of the four Dirac cones within the first BZ (Fig. 4(a)). We proceed to study the composition dependence of the $E^*$ peaks, and track the energy of the mass gap given by $\Delta=|E^*_+ - E^*_-|$. We find that $\Delta$ monotonically decreases from ~24.2 meV for *x*~0.38 to ~0meV near the critical doping of *x*~0.17 (Fig. 4(c)). Since this mass was attributed to a symmetry-breaking structural distortion, in theory, the mass gap may be controlled by changes in the magnitude of this distortion[11]. To establish the mechanism of the mass formation and its variation across the phase diagram it is therefore critical to pinpoint and directly image any distortion present in our samples.

In principle, STM images should be able to detect structural deformations. However, no distortion breaking mirror planes could be imaged in previous studies, since conventional low-bias topographs (Fig. 2(c)) showed only the Se sub-lattice. In the present study, by systematically adjusting the imaging conditions[19], we were able to detect both sublattices (Fig. 2(d)). At higher bias voltages, the Pb/Sn sublattice and the Se lattice are simultaneously observed. While a distortion can be observed in local patches by eye, to determine if a long-range phase coherent distortion exists in our samples, we create an average 4 atom by 4 atom ``supercell'' [20,21] (inset in Fig. 2(c) and Fig. 2(d)). We observe a characteristic zig-zag pattern associated with the relative in-plane atomic shift between the two sublattices (Fig. 2(a)) that signals the existence of a distortion in the surface layer (see *Supplementary Discussion I* for more details). This distortion, also known to occur within the bulk of the end member SnSe [22], breaks one of the two [110] mirror symmetries (Figs. 2(a,b)), leaving the other one unaffected as previously hypothesized [11]. To further substantiate the symmetry breaking observed in real-space, we examine the two-dimensional Fourier transforms (FTs) of the STM topographs (Figs. 2(e-j)). When only one sublattice is imaged at lower biases, the real parts of the Fourier components at the atomic Bragg peak wavevectors $Q_x$ and $Q_y$ show the same magnitude (Figs. 2(e,i)), while the respective imaginary parts are negligible (Fig. 2(g)). This observation demonstrates the $C_4$ symmetry of our STM tip. Next, in STM topographs showing both sublattices acquired using the same STM tip, we find that an imaginary component emerges along exclusively one lattice direction ($Q_y$ but not $Q_x$). This corresponds to the $C_4$ symmetry breaking in the sample crystal structure (Figs. 2(h,j)). Our results present the first direct evidence for the generation of massive Dirac fermions out of the pool of their massless counterparts due to a spontaneously broken crystal symmetry.

Next, we measure the magnitude of the distortion as a function of alloy composition to correlate it with the varying Dirac mass. Surprisingly, we find that the magnitude of the structural distortion does not significantly change across the phase diagram (Fig. 4(d)), which implies that the mass variation occurs due to a different process. Our data suggest an alternative explanation for the Dirac gap variation based on two factors: (1) the broken crystal symmetry is present exclusively on the surface, and (2) the surface state penetration depth $\lambda_{SS}$ increases on approaching $x_c$. In this scenario, the fraction of Dirac electrons sampling the broken symmetry region decreases on approaching $x_c$, effectively 'washing out' the effect of the surface symmetry breaking, and thus creating a smaller Dirac mass. To determine the validity of the scenario, we first determine whether the bulk is distorted. Our X-ray measurements in fact show a temperature independent, bulk cubic structure supporting the idea that the cubic symmetry is broken only at the surface (Fig. 1(b)).

Let us now consider the second factor. According to theory, the weight of Dirac SS at the surface of a topological insulator is not a constant. An estimate for length scale of the surface modes in z-direction (perpendicular to the surface) is: $\lambda_{SS} = \hbar v/E_g$, where $v$ and $E_g$ are the velocity and the gap of bulk Dirac fermions [23]. This implies that if one could continuously tune the bulk band gap, as the gap magnitude shrinks, the Dirac SS progressively penetrate further into the bulk, ultimately penetrating the entire sample at the critical point when the bulk gap closes. We note that this fundamental behavior of Dirac SS has however not yet been experimentally observed. To determine the actual weight of the SS as a function of composition in our samples, we use band structure parameters from our tight-binding model to calculate the weight of the topmost layer in the total SS wavefunction with varying Sn composition (Fig. 4(e)). Our calculations reveal that the surface contribution indeed decreases upon approaching $x_c$, and thus the penetration depth tends to infinity near the quantum critical point. Our experiments therefore indicate that the magnitude of the mass in our samples is controlled by the SS penetration into the bulk. Remarkably, our data represent the first observation of the striking effects of the divergent SS penetration depth on topological SS as the bulk approaches the critical concentration.

Having understood the mass formation, we now turn to the evolution of the SS dispersion across the critical composition, schematically portrayed in Fig. 5(a). To extract the dispersion of the SS bands, we first need to index field-dispersing LLs and plot them as a function of $(nB)^{0.5}$.

As previously demonstrated [11], this can be a nontrivial process due to the complex band structure of topological SS involving multiple Dirac cones which hybridize and go through a Lifshitz transition. We determine the LL indices by using the $(nB)^{0.5}$ scaling behavior to collapse the dispersion onto a continuous line. At the Lifshitz transition, area of the constant energy contour rapidly changes, which results in two van-Hove singularities (VHS) observed in our dI/dV spectra (Figs. 3(c,d)), and a "jump" in the LL indices (for example in Fig. 3(g), LL index jumps from n=1 to n=3). We begin with the LL plot of the topological sample at $x\sim0.34$ (Fig. 6(d)), which reveals many of the features of the SS band structure including the Lifshitz transitions schematically shown in Fig. 6(h). In the sample with lower Sn concentration (Fig. 6(c)), we find that both the Dirac points of the upper and lower Dirac cones and VHS within each pair move closer to each other in energy compared to that in the x~0.34 sample (Fig. 5(b) and Fig. 6(g)), consistent with previous ARPES studies on a closely related TCI family [24,25]. Following this trend, in the sample close to the transition point ($x\sim0.17$), the SS almost completely overlap each other (within our resolution, we see only one) as schematically depicted in Fig. 5(a). At this critical concentration, the V-shape of the dI/dV spectrum near the Dirac point disappears. Since the V-shape reflects the SS density of states within the bulk band gap, its absence suggests that the bulk gap at this concentration is close to zero, consistent with the critical point. Our data allows us to visualize the smooth evolution of the Lifshitz transition energy with varying alloy composition (Fig. 5(b)).

Finally, we turn our attention to the SS observed in the trivial phase below $x_c$ (Fig. 3(a,e)). Remarkably, utilizing the same method of scaling used for LL indexing of topological samples (Figs. 6(b-d)), we obtain a double-branched dispersion for both positive and negative LLs (Fig. 6(a)). Due to absence of a Dirac point and the double-branch dispersion, one might simply attribute these SS to conventional Rashba-split states, most notably known to occur on metal surfaces. However, the characteristics of the observed SS differ substantially from conventional Rashba states and should instead be viewed as topology induced proximity states. In contrast to trivial surface states in metals, such as those occurring on Cu(111) or Au(111) which have two quadratic dispersions shifted in momentum space [26,27], both branches of our proximity SS display linear dispersion away from the band bottom (coefficient of determination $R^2$ for the linear regression of all branches ~ 0.99) (Fig. 6(a)), with the dispersion velocity being constant (within ~7%) in all samples (Fig. 5(b)). This indicates an intimate connection between SS in the topological and non-topological regimes.

These observations strongly suggest that the persistent SS in trivial TCIs are remnants of the nearby topological phase, rather than a simple consequence of the boundary conditions at the surface. Physically, the trivial SS can be understood in the following way. Like the topological SS, we start with two Dirac cones at the X point; however, as the bulk transitions into the trivial regime, the mirror symmetry does not protect the SS from being fully gapped out (Fig. 6(e)) leading to the observed SS. We note that the splitting of the two branches is ~0.01 Å$^{-1}$, which is at the border of momentum-resolution of state-of-the-art laser ARPES studies, and could explain the inability to observe similar SS splitting [28,29]. Our findings open up a new venue for characterizing and engineering highly tunable two-dimensional electron gas with novel properties. The highly symmetric nature and linear dispersion of the proximity SS bands and the large gap could potentially be exploited in constructing a new generation of electronics, based on Dirac SS. Furthermore, the topological SS which harbor both massive and massless Dirac fermions present a unique platform for future applications, as they can be independently tuned via structural symmetry breaking and alloying composition.

## Methods

$Pb_{1-x}Sn_xSe$ single crystals with various Sn concentrations were grown by self-selecting vapor growth method, cleaved in ultra-high-vacuum at 77 K, and immediately inserted into the STM head. In our experiment, we determine the composition by performing the energy-dispersive X-ray spectroscopy (EDS) on multiple spots on each sample and averaging the obtained values.

# Figure 1

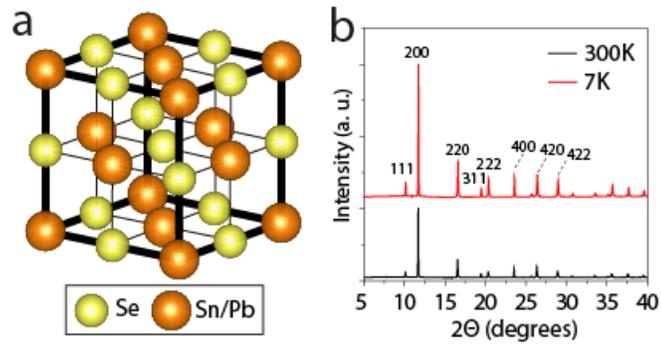

**Figure 1.** Bulk cubic crystal structure of $Pb_{1-x}Sn_xSe$. (a) Schematic of the face-centered cubic crystal structure showing the Se and Pb/Sn sublattices as yellow and orange spheres respectively. (b) X-ray diffraction intensity data for x~0.35 showing the cubic crystal structure without any distortion detected both at room temperature (black) and 7 K (red).

# Figure 2

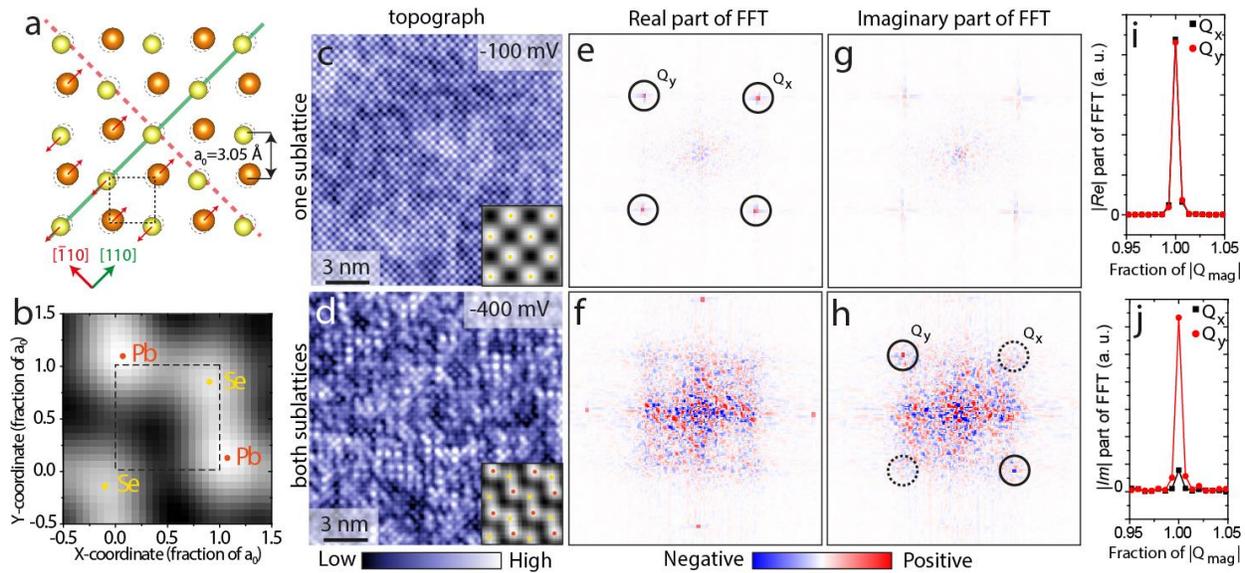

**Figure 2.** Symmetry-breaking distortion on the surface of $Pb_{1-x}Sn_xSe$. (a) Schematic representation of the distortion in the surface layer. Two sublattices shift in opposite manner along [110] direction as denoted by red arrows. Even though [110] mirror symmetry is still preserved (green line), the distortion breaks [$\bar{1}$10] mirror symmetry (dashed red line). Dashed black circles represent the ideal atomic positions in a cubic lattice without the distortion. (b) Average 2 atom by 2 atom supercell obtained from experimental topograph in (d). The centers of average atoms determined by Gaussian fitting are denoted by yellow and orange dots, and the dashed square represents the tetragonal unit cell in the absence of distortion (see *Supplementary Discussion I* for more details). STM topographs of x~0.38 sample showing (c) only Se sublattice (see *Supplementary Discussion II*), and (d) both sublattices acquired over the same area of the sample. Insets in (c,d) show 4 atom by 4 atom supercell aligned along the atomic Bragg peak directions [21]. Yellow and orange dots in insets in (c,d) represent centers of Se and Pb/Sn atoms respectively. (e,f) Real part of the Fourier-transform of STM topographs in (c,d). (g,h) Imaginary part of the Fourier-transform of STM topographs in (c,d). (i) Line cut starting from the center of FT in (e) through the atomic Bragg peaks $Q_x$ and $Q_y$ showing the $C_4$ symmetry of our STM tip. (j) Line cut starting from the center of FT in (h) through the atomic Bragg peaks $Q_x$ and $Q_y$ showing the symmetry breaking distortion present along only $Q_y$ direction. STM setup conditions are: (c) $I_{set}$=200pA, $V_{set}$=-100mV ; (d) $I_{set}$=900pA, $V_{set}$=-400mV.

**Figure 3**

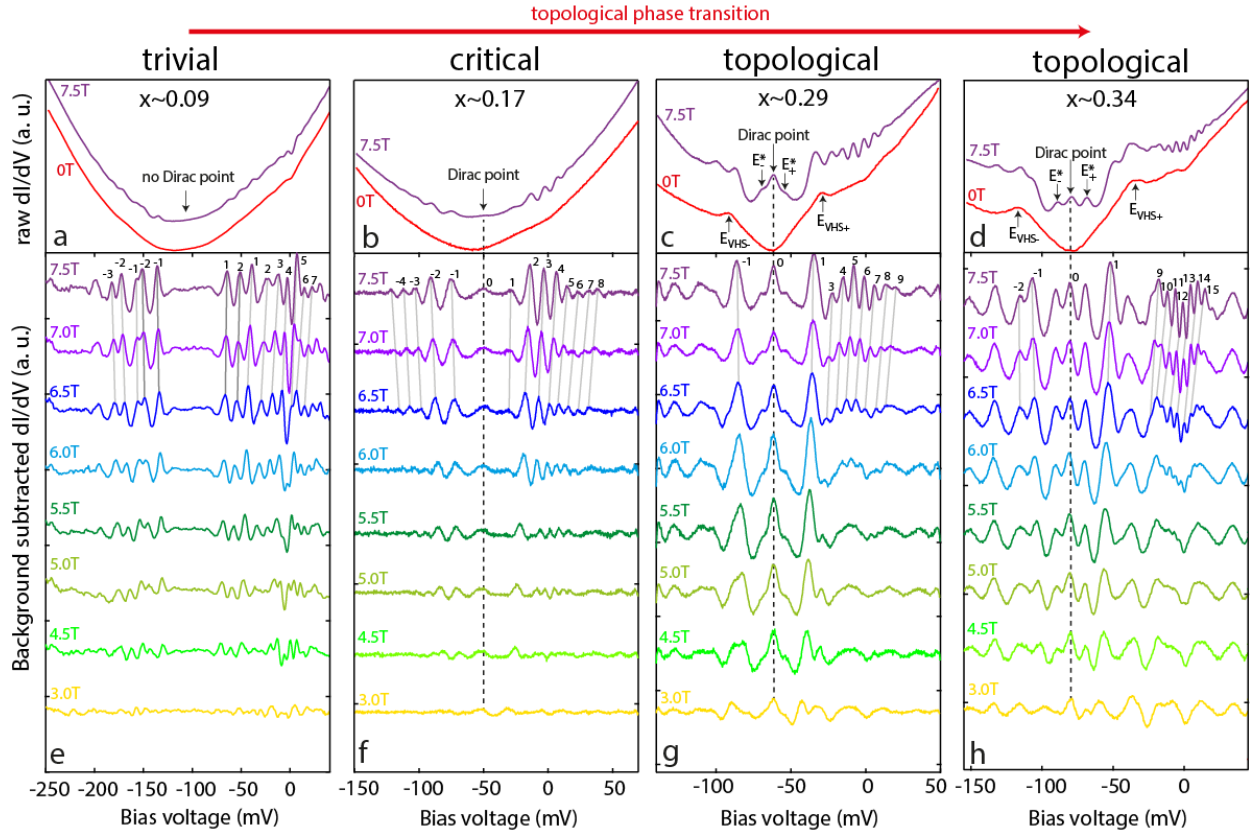

**Figure 3.** Landau level spectroscopy. (a-d) Average *dI/dV* curves acquired at 0T and 7.5T magnetic fields over the same area of the sample for x~0.09, 0.17, 0.29 and 0.34 Sn fraction, respectively. (e-h) Waterfall plots of background subtracted *dI/dV* curves (see *Supplementary Discussion III* for normalization details) acquired as a function of magnetic field, showing LLs as prominent dispersing ripples. Indexing of LLs used to plot the dispersions in Figure 6 is denoted in each panel. No Dirac point is present in the trivial sample (e), while a prominent non-dispersing peak which develops under the magnetic field application signals the Dirac node formation in (f-h). Other non-dispersing peaks in (c,d,g,h) labeled as $E^*_+$ and $E^*_-$ indicate the presence of massive Dirac fermions [11]. $E_{VHS+}$ and $E_{VHS-}$ denote the positions of two van-Hove singularities, which are directly related to the two Lifshitz transitions. STM setup conditions are: (a) $I_{set}$=300pA , $V_{set}$=-200mV ; (b) $I_{set}$=385pA , $V_{set}$=70mV ; (c) $I_{set}$= 540pA, $V_{set}$=50mV ; (d) $I_{set}$=350pA , $V_{set}$=50mV .

**Figure 4**

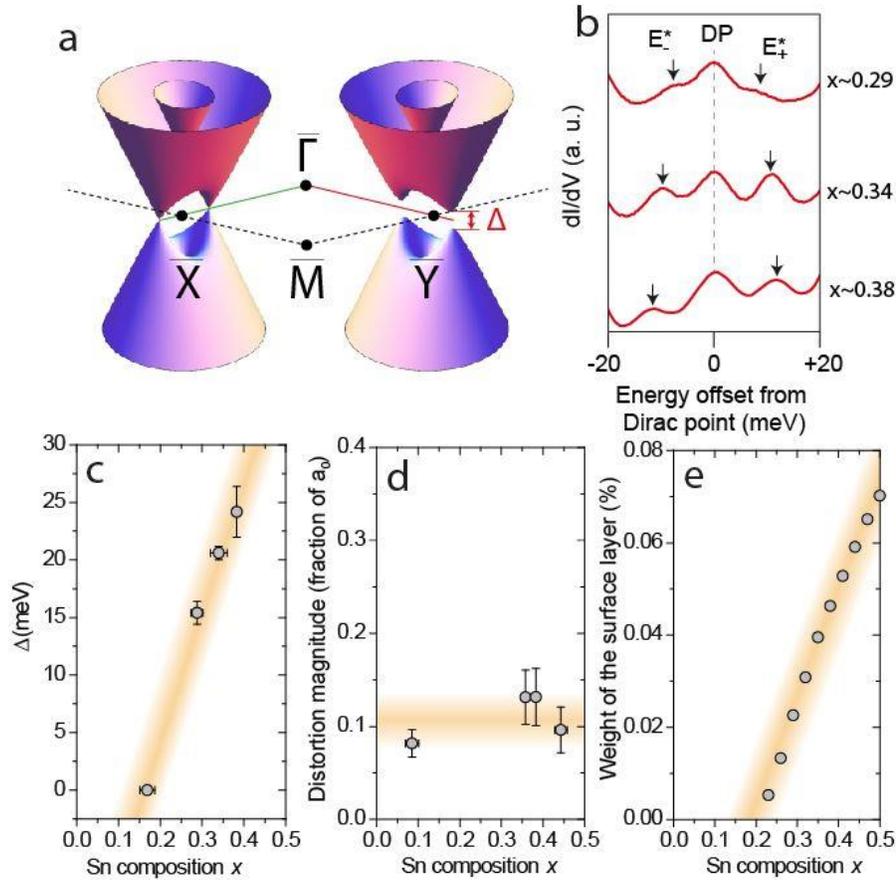

**Figure 4.** Evolution of the Dirac gap with alloy composition. (a) Illustration of the Dirac gap Δ opening in two out of the four Dirac cones within the first BZ due to a broken [110] mirror plane. (b) Composition dependence of average *dI/dV* spectra acquired at 7.5T, showing the evolution of the non-dispersing E* LL peaks. (c) Plot of Δ vs. Sn composition showing the monotonic increase in Δ in the topological regime. Orange line in (c) represents the linear fit to the experimental data points. (d) Plot of distortion magnitude as a function of Sn composition showing the presence of distortion of similar magnitude across the wide range of compositions (see *Supplementary Discussion IV* for more details). (e) Theoretical calculations of the surface layer contribution to the Dirac wavefunction showing a monotonic increase (penetration depth effectively decreases) with increased Sn doping. Horizontal error bars in (c,d) denote the standard deviation in EDS measurements of Sn composition from 5 different areas of each sample. Vertical error bars in (c,d) are explained in *Supplementary Discussion III* and *IV*, respectively.

**Figure 5**

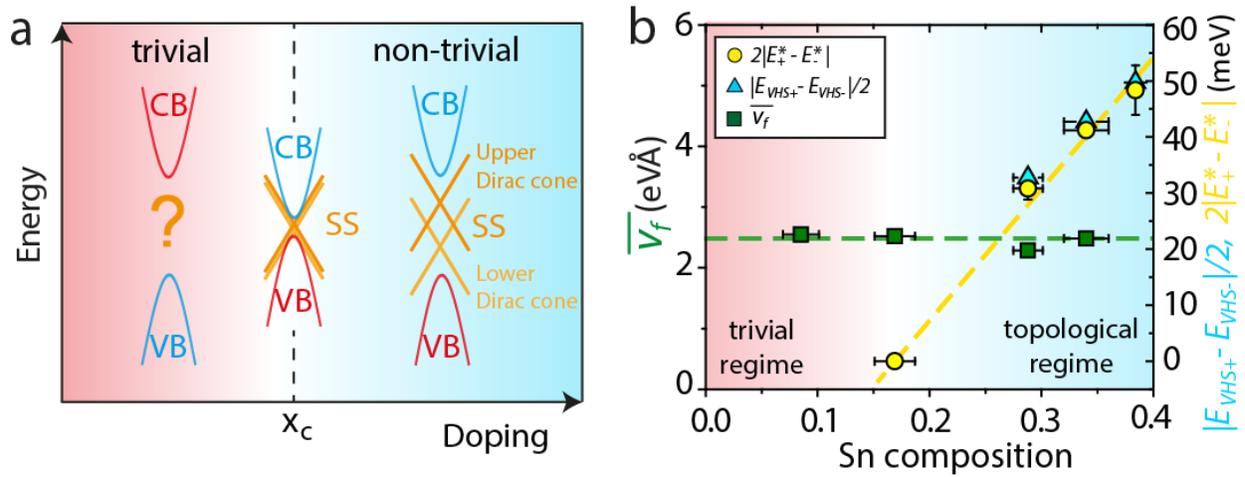

**Figure 5.** Band structure evolution with alloy composition. (a) Schematic phase diagram of TCIs. The band gap between the conduction (CB) and valence bands (VB) closes at a critical Sn concentration $x_c$ of approximately 0.18 [12], causing band inversion and the formation of topologically-protected SS consisting of two Dirac cones centered at each X point. (b) Evolution of band structure parameters as a function of Sn composition. Average SS dispersion velocity $v_f$ assuming isotropic momentum dispersion, 2Δ and the energy of the Lifshitz transition are shown as green squares, yellow circles and blue triangles, respectively. Horizontal error bars in (b) denote the standard deviation in EDS measurements of Sn composition from 5 different areas of each sample. Vertical error bars in (b) are explained in *Supplementary Discussion III*.

**Figure 6**

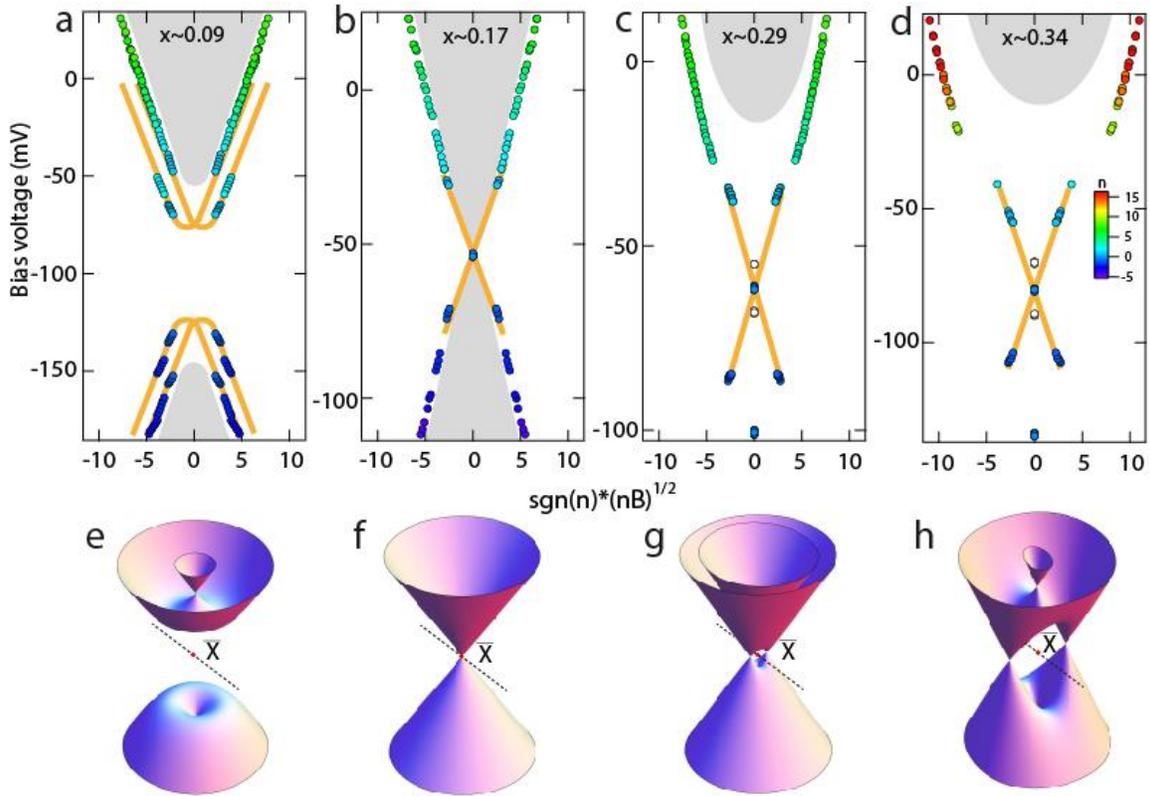

**Figure 6.** SS evolution across the topological quantum phase transition. Plots of LL dispersion (a) before, (b) at, and (c,d) after the topological transition. The double-branch dispersion in (a) shows the proximity state without the Dirac node, which is clearly observed in topological samples in (b-d). Two Dirac cones are nearly indistinguishable in (b), but become increasingly offset in energy upon increasing Sn content (c,d). Non-dispersing E* peaks in (c,d) (white circles), are absent in the transition sample in (b), which allows us to visualize the effect of composition changes on the Dirac gap, a property directly related to the Dirac fermion mass. Orange lines in (a-d) schematically depict the low energy SS dispersions. Gray shaded regions portray the positions of the bulk bands determined from Ref.[12]. Fits to the ***k.p*** model [30] are shown in *Supplementary Discussion V*. (e-h) Illustrations of the SS band structure around a single X point in the absence of symmetry breaking surface distortion, showing its evolution from trivial to the topological regime. The lack of separate $E^*_+$ and $E^*_-$ peaks at the transition point x~0.17 (Fig. 3(b)) suggests that Δ is approximately zero at this concentration.